\documentclass[11pt,fleqn]{article}
\usepackage{amssymb}
\newcommand{\Section}[1]%
{\section{#1}\setcounter{equation}{0}%
\setcounter{theorem}{0}}

{\par\noindent{\em #1:\ }}%
{~\rule{2mm}{2mm}\par\bigskip}
\def\define{\stackrel{{\rm def}}{=}}
\newcommand{\ret}{\nonumber \\}
\newcommand{\beq}{\begin{equation}}
\newcommand{\eeq}{\end{equation}}
\newcommand{\bem}{\begin{displaymath}}
\newcommand{\eem}{\end{displaymath}}
\newcommand{\beqar}{\begin{eqnarray}}
\newcommand{\eeqar}{\end{eqnarray}}

\newcommand{\abs}[1]{\left| #1 \right|}

\newcommand{\dagg}{{\scriptscriptstyle\dagger}}
\font\titlefnt=cmbx10 scaled \magstep2

\begin{document}
\mathindent 0mm
\newpage\thispagestyle{empty}
\begin{flushright} HD--TVP--96--9, cond--mat/9609065 \end{flushright}
\vspace*{2cm}
\begin{center}
{\titlefnt Similarity renormalization of the
electron--phonon coupling\\ \vspace*{0.4cm}
}
\vskip3cm
Andreas Mielke\footnote[1]
{E--mail: mielke@hybrid.tphys.uni-heidelberg.de}\\
\vspace*{0.2cm}
Institut f\"ur Theoretische Physik,\\
Ruprecht--Karls--Universit\"at,\\
Philosophenweg 19, \\
D-69120~Heidelberg, F.R.~Germany
\\
\vspace*{1.5cm}
revised  version, \today
\\
\vspace*{2cm}
\noindent

{\bf Abstract}
\end{center}

\vspace*{0.2cm}\noindent
We study the problem of the phonon--induced electron--electron
interaction in a solid. Starting with a Hamiltonian that contains
an electron--phonon interaction, we perform a similarity renormalization
transformation to calculate an effective Hamiltonian.
Using this transformation singularities due to degeneracies
are avoided explicitely.
The effective interactions are calculated to second order
in the electron--phonon coupling. It is shown that the
effective interaction between two electrons forming a
Cooper pair is attractive in the whole parameter space.
For a simple Einstein model we calculate the renormalization of
the electronic energies and the critical temperature of
superconductivity.

\vspace*{1cm}
Keywords: Electron--phonon coupling, effective interactions,
superconductivity

\vspace*{2cm}
\newpage
\topskip 0cm
\Section{Introduction}
BCS--theory \cite{BCS} is one of the most successful and most popular
theories in condensed matter physics. It
is based on a Hamiltonian with an attractive interaction
between the electrons. Phonons are not explicitely
present in this model. The phonon frequencies enter
in the explicit form of the phonon--induced effective
electron--electron interaction. Most of the properties
of superconductors can be understood using BCS--theory. But
one of the problems of BCS--theory is that the critical
temperature $T_c$ of the superconductor cannot be
calculated. The famous BCS--formula
$T_c=1.13\Theta \exp(-1/N(\epsilon_F)V)$,
where $\Theta$ is the Debye temperature and $V$ is
the strength of the interaction, has to
be used to determine $N(\epsilon_F)V$ from $T_c$.
If one uses for $N(\epsilon_F)V$ values that have been
obtained using standard perturbative treatments
of the electron--phonon interaction,
the calculated critical temperature is too large.
Reliable results for $T_c$ can be obtained in the framework
of the Eliashberg theory \cite{Eliashberg}. For a review
of this theory and a critical discussion of various formulas
for the critical temperature we refer to \cite{Scalapino, Allen82},
see also \cite{Allen90}.

In their discussion of BCS--theory, Allen et al \cite{Allen82}
argued that the BCS--equation yields a wrong result for $T_c$ because
retardation effects in the interaction are neglected.
This point of view is possible in a formulation of the
theory in terms of Green's functions like in the Eliashberg--theory
\cite{Allen82} or equivalently in a field--theoretic formulation
of a model containing electrons and phonons,
where it is possible to integrate out the phonons.
In both cases one obtains an effective phonon--induced
electron--electron interaction that depends on time
or frequency. But in the standard formulation of BCS--theory
one starts with a Hamiltonian. The effective electron--electron
interaction Bardeen et al \cite{BCS} had in mind was the
phonon--induced interaction of Fr{\"o}hlich \cite{Froehlich}
or Bardeen and Pines \cite{BP}. In these approaches the
phonon--induced interaction is obtained by applying a
unitary transformation to eliminate the electron--phonon
interaction in the Hamiltonian in lowest order. It is clear that
in a Hamiltonian formulation of the theory, the interaction
cannot be frequency dependent, since the Hamiltonian is a hermitian
operator. From the viewpoint of a Hamiltonian formulation the
problem of BCS--theory with an interaction of the Fr{\"o}hlich type
is that the interaction is singular and
does not contain the correct energy scale.
If one performs a single unitary transformation to eliminate
the electron--phonon interaction, one attempts to treat all
energy scales in the problem at once. This usually fails, even
in perturbation theory. Instead one should break the problem into pieces,
dealing with each energy scale in sequence. This is the
usual approach of renormalization theory.

About two years ago G{\l}azek and Wilson \cite{Glazek94} proposed a new
renormalization scheme for Hamiltonians, called {\em
similarity renormalization}. The main idea of this approach
is to perform a continuous unitary transformation that
yields a band--diagonal effective Hamiltonian.
In the effective Hamiltonian the dependence on
the original ultra--violet cutoff is removed,
but it will contain additional interactions.
In the present paper we apply this renormalization
scheme to the classical problem of interacting electrons
and phonons, modelled by a Hamiltonian
\beq
H=\sum_k\epsilon_{k}:c^\dagg_k c_k:
+\sum_q\omega_{q} :b^\dagg_q b_q:
+\sum_{k,q}(g_{q}c^\dagg_k c_{k+q} b^\dagg_q
+g^*_{q}c^\dagg_{k+q} c_k b_q).
\eeq
$c^\dagg_k$ and $c_k$ are the usual creation
and annihilation operators for electrons,
$b^\dagg_q$ and $b_q$ for phonons, and the colons
denote normal ordering.
We calculate the renormalized Hamiltonian
to second order in the coupling constant $g_q$ of the
electron--phonon interaction. It contains an
additional effective electron--electron interaction,
that is responsible for superconductivity.

Recently, the construction of the effective
phonon--induced electron--electron interaction has been studied by Lenz and
Wegner \cite{Lenz96}. They used flow equations for Hamiltonians,
a method  proposed by
Wegner \cite{Wegner94} to block--diagonalize a given
Hamiltonian. They obtained an expression for the
effective interaction that differs from Fr{\"o}hlich's result
\cite{Froehlich},
in fact it is less singular.
If we write the induced electron--electron interaction
in the form
\beq
\frac12\sum_{k,k^\prime,q}V_{k,k^\prime, q}
:c^\dagg_{k+q} c^\dagg_{k^\prime-q} c_{k^\prime} c_{k}:,
\eeq
the coefficient for the interaction of two
electrons forming a Cooper pair obtained by Fr{\"o}hlich is
\beq \label{Fint}
V_{k,-k, q}=\abs{g_q}^2\frac{2\omega_q}
{(\epsilon_{k+q}-\epsilon_k)^2-\omega_q^2}
\eeq
whereas Lenz and Wegner obtained
\beq \label{LWint}
V_{k,-k, q}=-\abs{g_q}^2\frac{2\omega_q}
{(\epsilon_{k+q}-\epsilon_k)^2+\omega_q^2}.
\eeq
There are two remarkable differences between these results.
First, the interaction obtained by Lenz and Wegner
has no singularity, second it is attractive for
all $k$ and $q$.

Recently Wegner's flow equations for Hamiltonians have been
applied to the Anderson impurity model \cite{KM}. For that model
a similar problem occurs. It can be mapped onto
the Kondo model using the well known
Schrieffer--Wolff transformation \cite{SW}.
The Schrieffer--Wolff transformation eliminates the hybridization
between the impurity state and the electronic band states
to first order. Using this transformation, the induced spin--spin
interaction contains a singularity if the impurity
orbital lies in the electronic band. Furthermore,
the induced interaction does not have the
correct energy scale built in, therefore the Kondo
temperature can not be calculated correctly. The flow
equations yield a less singular expressions which
has a different sign in some regions of the parameter space.
In the case of the Anderson impurity model it was possible
to compare the results with renormalization group calculations,
and it could be shown that the flow equations yield the correct
behaviour. For instance, the energy scale which sets the scale for the
Kondo temperature comes out correctly, whereas the
Schrieffer--Wolff result does not have this property \cite{KM}.

Unfortunately, a similar comparison with renormalization
group calculations was not possible in the case
of the electron--phonon coupling.
The primary purpose of the present paper is to fill this gap.
Using the above mentioned similarity renormalization scheme,
we will calculate the effective electron--electron interaction.
We will show that the interaction between two electrons
forming a Cooper pair has the sign structure of
the expression obtained by Lenz and Wegner,
it is attractive for all $k$ and $q$.
This clearly supports the result by Lenz and Wegner (\ref{LWint}).
The difficulties of the $T_c$ formula of BCS--theory
can be traced back to the wrong effective
electron--electron interaction. We will come back to this
point in Sect. 4.

For a reader who is familiar with Eliashberg theory,
it is probably unclear how an approach using a Hamiltonian
framework can lead to quantitative results without a fit
parameter like in BCS--theory. Neither the interactions
(\ref{Fint}) or (\ref{LWint}) nor the interaction obtained
using similarity renormalization contain retardation effects.
The reason that the procedure yields reliable results
is that similarity renormalization yields
the correct low energy scale of the problem.
Different energy scales are  automatically separated.
How this happens will be made explicit in Sect. 2.
A second point is that since the renormalization procedure
consists of a continuous unitary transformation applied
to the Hamiltonian, one has to transform the observables as well.
Retardation, i.e. the decay of single particle states
and the broadening of peaks in spectral functions can be
obtained within the present approach, if the transformation of
the observables is taken into account. In a recent paper
on time--dependent equilibrium correlation functions in dissipative
quantum systems it has been demonstrated, how the transformation
of the observables can be done \cite{KM2}. In that case Wegner's flow
equation \cite{Wegner94} have been used, but the transformation
of the observables is very similar in both approaches. It was
possible to obtain very accurate quantitative results in that case.
In the present paper we only calculate static properties
of the system, therefore we will not discuss this point further.

The outline of this paper is as follows.
In the next section we give a brief description
of the similarity renormalization scheme. For details
we refer to \cite{Glazek94}. This method has been
applied to quantum electrodynamics \cite{Perry1}
and to quantum chromo dynamics \cite{Wilson94, Perry2}
(and the references therein). Many details
concerning the method are explained in these papers
as well.
In Sect. 3 we apply the method to the
electron--phonon coupling. We derive
the renormalization group equations and calculate the renormalized
Hamiltonian to second order in $g_q$.
Mainly for illustrational purpose we derive some explicit results
for the simple Einstein model for phonons in Sect. 4. First we
calculate the renormalization of the electronic single particle
energies. They show a typical logarithmic renormalization,
which leads to the well known renormalization of the
density of states at the Fermi surface.
Next we calculate $T_c$ within the Einstein model
using our effective interaction. We compare our results with those
obtained using (\ref{LWint}) or using the
$T_c$--formula of BCS-theory. Finally we give some conclusions including
a comparison between the similarity renormalization
scheme and Wegner's flow equations.
\Section{Similarity renormalization}
This method yields renormalization equations for
a given arbitrary Hamiltonian $H_\lambda$ as a function of
the cutoff $\lambda$. The goal of the
method is to transform the initial Hamiltonian
(with a large cutoff $\Lambda$) into an effective
Hamiltonian that has no matrix elements
providing energy jumps large compared to
the small cutoff $\lambda$.
The initial Hamiltonian contains usually multiple energy
scales and couplings between these energy scales.
The renormalized Hamiltonian $H_\lambda$ has a band diagonal form.
The different energy scales in the initial
Hamiltonian are decoupled in $H_\lambda$.

Let us write $H_\lambda$ in the form
\beq
H_\lambda=H_{0\lambda}+H_{I\lambda}.
\eeq
$H_{0\lambda}$ is the free Hamiltonian, $H_{I\lambda}$ contains
interactions and counter terms. The eigenvalues of $H_{0\lambda}$
are $E_{i\lambda}\ge 0$. We introduce
cutoff functions $u_{ij\lambda}$ for the matrix elements.
To assure a band diagonal structure of the renormalized
Hamiltonian, we choose $u_{ij\lambda}=1$ if
$\abs{E_{i\lambda}-E_{j\lambda}}$ is small compared
to lambda and $u_{ij\lambda}=0$ if
$\abs{E_{i\lambda}-E_{j\lambda}}$ is large compared to
$\lambda$. The detailed form of $u_{ij\lambda}$
is not important. One possible choice, which has
been used in the treatment of QED on the light front
\cite{Perry1} is
\beq \label{sharp_cutoff}
u_{ij\lambda}=\theta(\lambda-\abs{E_{i\lambda}-E_{j\lambda}}).
\eeq
In other situations one has to introduce a
smooth cutoff function. For a discussion of this point
we refer to \cite{Glazek94, Wilson94}.
Furthermore we introduce
\beq
r_{ij\lambda}=1-u_{ij\lambda}
\eeq
The renormalization of $H_\lambda$ can be written down
as an infinitesimal unitary transformation
\beq
\frac{dH_\lambda}{d\lambda}=[\eta_\lambda, H_\lambda]
\eeq
with a generator $\eta_\lambda$. $\eta_\lambda$
has to be chosen so that the matrix elements of
$H_\lambda$ obey
\beq \label{sr0}
H_{ij\lambda}=u_{ij\lambda}Q_{ij\lambda}.
\eeq
This assures that $H_\lambda$ is band diagonal.
For the matrix elements $Q_{ij\lambda}$ we obtain
\beq
\frac{du_{ij\lambda}}{d\lambda}Q_{ij\lambda}
+u_{ij\lambda}\frac{dQ_{ij\lambda}}{d\lambda}
=\eta_{ij\lambda}(E_{i\lambda}-E_{j\lambda})
+[\eta_\lambda,H_{I\lambda}]_{ij}.
\eeq
One defines
\beq
G_{ij\lambda}\define
[\eta_\lambda,H_{I\lambda}]_{ij}
-\frac{du_{ij\lambda}}{d\lambda}Q_{ij\lambda}
\eeq
and chooses
\beq
\eta_{ij\lambda}(E_{i\lambda}-E_{j\lambda})
=-r_{ij\lambda}G_{ij\lambda}.
\eeq
This yields the final equations for  the matrix elements of the
generator of the infinitesimal unitary transformation
\beq \label{sr1}
\eta_{ij\lambda}=\frac{r_{ij\lambda}}{E_{i\lambda}-E_{j\lambda}}
\left([\eta_\lambda,H_{I\lambda}]_{ij}-
\frac{du_{ij\lambda}}{d\lambda}\frac{H_{ij\lambda}}{u_{ij\lambda}}\right),
\eeq
and for the renormalization
of the Hamiltonian
\beq \label{sr2}
\frac{dH_{ij\lambda}}{d\lambda}
=u_{ij\lambda}[\eta_\lambda,H_{I\lambda}]_{ij}
+r_{ij\lambda}\frac{du_{ij\lambda}}{d\lambda}
\frac{H_{ij\lambda}}{u_{ij\lambda}}.
\eeq
Since $r_{ij\lambda}=0$ if
$\abs{E_{i\lambda}-E_{j\lambda}}\lessapprox\lambda$,
the denominator $E_{i\lambda}-E_{j\lambda}$ in (\ref{sr1}) is
bounded from below.
These equations are exact, but
it is clear that one is not able to solve them
explicitely. Suitable approximations are
necessary. Since small energy denominators are
avoided in this approach one possibility is
a systematic expansion in a coupling constant.

Before we proceed to the application of the similarity
renormalization, let us take a closer look at the
equations (\ref{sr1}) and (\ref{sr2}). The main point
is that due to (\ref{sr0})
the renormalized Hamiltonian contains only couplings
between states with an energy difference less than $\lambda$.
Starting with an initial cutoff $\Lambda$, couplings between
very different energy scales are eliminated first, couplings
between states belonging to the same energy scale are
not eliminated or are eliminated later for smaller values of $\lambda$.
As a consequence, different energy scales are seperated during the
renormalization process.
\Section{Application to the electron--phonon coupling}
The Hamiltonian of the model is
\beq
H_\lambda=H_{0\lambda}+H_{I\lambda}
\eeq
with
\beq
H_{0\lambda}=\sum_k\epsilon_{k,\lambda}:c^\dagg_k c_k:
+\sum_q\omega_{q,\lambda} :b^\dagg_q b_q:,
\eeq
\beq
H_{I\lambda}=\sum_{k,q}:g_{k,q,\lambda}c^\dagg_k c_{k+q} b^\dagg_q
+g^*_{k,q,\lambda}c^\dagg_{k+q} c_k b_q: +O(g^2).
\eeq
$c^\dagg_k$ and $c_k$ are the electron creation and annihilation
operators. The index $k$ is a shorthand notation for $({\bf k},\sigma)$
and $-k$ denotes $(-{\bf k},\sigma)$. $b^\dagg_q$ and $b_q$ are the
creation and annihilation operators for phonons. Let us
assume $\omega_{q,\lambda}=\omega_{-q,\lambda}$,
$\epsilon_{k,\lambda}=\epsilon_{-k,\lambda}$.
For a large cutoff $\Lambda$ we let
$g_{k,q,\Lambda}=g_{q,\Lambda}=g_{-q,\Lambda}$ independent of $k$.
The terms $O(g^2)$ in $H_{I\lambda}$ contain induced
interactions. For simplicity we do not include
a Coulomb repulsion in our model Hamiltonian.
We have introduced a normal ordering for the fermions and for the bosons,
\beq
:c^\dagg_k c_k:=c^\dagg_k c_k-n_k,
\eeq\beq
:b^\dagg_q b_q:=b^\dagg_q b_q-\bar{n}_q,
\eeq
where $n_k$ is the occupation number for electrons in the state
$k$ and $\bar{n}_q$ is the occupation number for the bosons, respectively.
The Fermi energy is set to zero.

The model contains several energy scales. The largest energy
scale is given by the band width of the electronic
single particle states, which is usually a few eV.
Typical phonon energies are given by the Debye frequency,
they are about two orders of magnitude smaller. Due to the
phonon induced interaction between the electrons, Cooper
pairs are formed. Their energy scale is given by the
critical temperature of superconductivity, which is again
about two orders of magnitude smaller than typical phonon energies.
One has to resolve an energy scale in a Hamilonian
that is dominated by energies being four orders of magnitude larger.
Such a situation often occurs when a given problem has a marginal relevant
operator. In our case this is the attractive phonon--induced
electron--electron interaction. When this interaction is
calculated within a perturbative scheme like the one
used by Fr{\"o}hlich, all energy scales are treated at once.
This usually fails.

We now investigate this model using (\ref{sr1}, \ref{sr2}).
All quantities will be calculated as series expansions in $g$.
Due to (\ref{sr1}) the generator of the infinitesimal
unitary transformation can be written as
\beq
\eta_\lambda=\sum_{k,q}:\eta_{k,q,\lambda}c^\dagg_k c_{k+q} b^\dagg_q
-\eta^*_{k,q,\lambda}c^\dagg_{k+q} c_k b_q: +O(g^2)
\eeq
with $\eta_{k,q,\lambda}=O(g)$.
In order to have a definite choice for
$u_{k,q,\lambda}$, we let
\beq
u_{k,q,\lambda}=u(
\abs{\epsilon_{k,\lambda}-\epsilon_{k+q,\lambda}+\omega_{q,\lambda}}
/\lambda)
\eeq
with a smooth cutoff function $u(x)$ that drops fast from 1 to
zero in the vicinity of $x=1$.
Then (\ref{sr1}) yields
\beq \label{sr-eta1}
\eta_{k,q,\lambda}=-\frac{r_{k,q,\lambda}}
{\epsilon_{k,\lambda}-\epsilon_{k+q,\lambda}+\omega_{q,\lambda}}
\frac{d\ln u_{k,q,\lambda}}{d\lambda}
g_{k,q,\lambda} + O(g^3).
\eeq
Let us now calculate the effective Hamiltonian.
The first term in (\ref{sr2}) is of order $g^2$.
This term contains the renormalization of
the single particle energies
$\epsilon_{k,\lambda}$ and $\omega_{q,\lambda}$
and some new, induced interaction  in $H_{I\lambda}$.
The second term in (\ref{sr2}) is of order $g$ and contributes
to the renormalization flow of the electron--phonon
coupling. The renormalization of this coupling is
determined by the equation
\beq
\frac{dg_{k,q,\lambda}}{d\lambda}=
r_{k,q,\lambda}\frac{d\ln u_{k,q,\lambda}}{d\lambda}
g_{k,q,\lambda}+O(g^3).
\eeq
It can be solved using the {\it ansatz}
\beq
g_{k,q,\lambda}=g_{q,\Lambda}
\frac{e_{k,q,\lambda}}{e_{k,q,\Lambda}}
\eeq
with $e_{k,q,\lambda}
=e(\abs{\epsilon_{k,\lambda}-\epsilon_{k+q,\lambda}+\omega_{q,\lambda}}
/\lambda)$.
$e(x)$ obeys
$\frac{1}{e}\frac{de}{dx}=\frac{r}{u}\frac{du}{dx}$.
Using $r=1-u$ it can be written in the form
\beq
e(x)=u(x)\exp(r(x)).
\eeq
$e(x)$ has properties similar to $u(x)$:
$e(x)=1$ if $u(x)=1$, $e(x)=0$ if $u(x)=0$,
$e(x)$ decays monotonously.
Let us assume that the cutoff $\Lambda$
is sufficiently large, so that, $e_{k,q,\Lambda}=1$
holds for all $k$, $q$. This yields
$g_{k,q,\lambda}=g_{q,\Lambda}e_{k,q,\lambda}$.

Let us now calculate the terms of order $g^2$.
To do this we have to calculate the commutator
$[\eta_\lambda,H_{I\lambda}]$. It is given by
\beqar
[\eta_\lambda,H_{I\lambda}]&=&
\sum_{k,q}\sum_{k^\prime , q^\prime}
\eta_{k,q,\lambda}
(g_{k^\prime, q^\prime, \lambda}[c^\dagg_k c_{k+q}b^\dagg_q,
c^\dagg_{k^\prime} c_{k^\prime+q^\prime}b^\dagg_{q^\prime}]
+g^*_{k^\prime ,q^\prime ,\lambda}[c^\dagg_k c_{k+q}b^\dagg_q,
c^\dagg_{k^\prime+q^\prime}c_{k^\prime} b_{q^\prime}])
\ret & & \quad
+\mbox{ h.c. } +\, O(g^3)
\ret &=&
\sum_{k,q}\sum_{k^\prime , q^\prime}
\eta_{k,q,\lambda}g_{k^\prime ,q^\prime ,\lambda}
(:c^\dagg_k c_{k+q+q^\prime}b^\dagg_q b^\dagg_{q^\prime}:
\delta_{k^\prime,k+q}
-:c^\dagg_{k-q^\prime} c_{k+q}b^\dagg_q b^\dagg_{q^\prime}:
\delta_{k^\prime,k-q^\prime}
\ret & & \qquad
+(n_k-n_{k+q}):b^{\dagg}_qb^{\dagg}_{-q}:
\delta_{q,-q^\prime}\delta_{k^\prime,k+q})
\ret &+&
\sum_{k,q}\sum_{k^\prime , q^\prime}
\eta_{k,q,\lambda}g^*_{k^\prime ,q^\prime ,\lambda}
(:c^\dagg_k c_{k+q-q^\prime}b^\dagg_q b_{q^\prime}:
\delta_{k^\prime,k+q-q^\prime}
-:c^\dagg_{k+q^\prime}c_{k+q}b^\dagg_q b_{q^\prime}:
\delta_{k^\prime,k}
\ret & & \qquad
+(n_k-n_{k+q}):b^{\dagg}_qb_q:
\delta_{q,q^\prime}\delta_{k^\prime,k}
\ret & & \qquad
-:c^\dagg_k c^\dagg_{k^\prime+q} c_{k^\prime} c_{k+q}:\delta_{q,q^\prime}
\ret & & \qquad
-(1-n_k+\bar{n}_q):c^\dagg_{k+q}c_{k+q}:\delta_{q,q^\prime}\delta_{k^\prime,k}
+(n_{k+q}+\bar{n}_q):c^\dagg_k c_k:\delta_{q,q^\prime}\delta_{k^\prime,k}
\ret & & \qquad
+((n_k-n_{k+q})\bar{n}_q
-(1-n_k)n_{k+q})\delta_{q,q^\prime}\delta_{k^\prime,k})
+\mbox{ h.c. }+\, O(g^3)
\eeqar
This expression together with (\ref{sr2}) yields
the renormalization of $\omega_{q,\lambda}$ and $\epsilon_{k,\lambda}$.
For $\omega_{q,\lambda}$ we obtain
\beq
\frac{d\omega_{q,\lambda}}{d\lambda}
=\sum_k(\eta_{k,q,\lambda}g^*_{k,q,\lambda}
+\eta^*_{k,q,\lambda}g_{k,q,\lambda})
(n_k-n_{k+q})
\eeq
Using (\ref{sr-eta1}) and the definition of $e_{k,q,\lambda}$
yields
\beq
\eta_{k,q,\lambda}=-\frac{g_{k,q,\lambda}}
{\epsilon_{k,\lambda}-\epsilon_{k+q,\lambda}+\omega_{q,\lambda}}
\frac{d\ln e_{k,q,\lambda}}{d\lambda}
\eeq
and therefore
\beqar
\frac{d\omega_{q,\lambda}}{d\lambda}
&=&\sum_k\frac{n_{k+q}-n_k}
{\epsilon_{k,\lambda}-\epsilon_{k+q,\lambda}+\omega_{q,\lambda}}
\frac{d\ln e^2_{k,q,\lambda}}{d\lambda}
\abs{g_{k,q,\lambda}}^2
\ret &=&
\sum_k\frac{n_{k+q}-n_k}
{\epsilon_{k,\lambda}-\epsilon_{k+q,\lambda}+\omega_{q,\lambda}}
\frac{d\abs{g_{k,q,\lambda}}^2}{d\lambda}.
\eeqar
We integrate this flow equation from a small cutoff $\lambda$
to the initial large cutoff $\Lambda$. On the right hand side
we perform an integration by parts. Then the remaining integral
contains derivatives of the single particle energies and is
therefore of fourth order in $g_{k,q,\lambda}$. Neglecting
this term we obtain
\beq \label{selfcon-omega}
\omega_{q,\lambda}=\omega_{q,\Lambda}
+\sum_k\frac{n_k-n_{k+q}}
{\epsilon_{k,\lambda}-\epsilon_{k+q,\lambda}+\omega_{q,\lambda}}
\abs{g_{q,\Lambda}}^2(1-e^2_{k,q,\lambda}).
\eeq
A similar calculation can be performed for $\epsilon_{k,\lambda}$.
The renormalization of $\epsilon_{k,\lambda}$ is determined by
\beqar
\frac{d\epsilon_{k,\lambda}}{d\lambda}&=&
\sum_q(\eta_{k,q,\lambda}g^*_{k,q,\lambda}(n_{k+q}+\bar{n}_q)+
\eta^*_{k,q,\lambda}g_{k,q,\lambda}(n_{k+q}+\bar{n}_q)
\ret & & \quad
-\eta_{k+q,-q,\lambda}g^*_{k+q,-q,\lambda}(1-n_{k+q}+\bar{n}_q)
-\eta^*_{k+q,-q,\lambda}g_{k+q,-q,\lambda}(1-n_{k+q}+\bar{n}_q))
\ret &=&
-\sum_q\frac{n_{k+q}+\bar{n}_q}
{\epsilon_{k,\lambda}-\epsilon_{k+q,\lambda}+\omega_{q,\lambda}}
\frac{d\abs{g_{k,q,\lambda}}^2}{d\lambda}
+\sum_q\frac{1-n_{k+q}+\bar{n}_q}
{\epsilon_{k+q,\lambda}-\epsilon_{k,\lambda}+\omega_{q,\lambda}}
\frac{d\abs{g_{k+q,-q,\lambda}}^2}{d\lambda}
\eeqar
and with the same assumptions as above we obtain
\beqar \label{selfcon-epsilon}
\epsilon_{k,\lambda}&=&\epsilon_{k,\Lambda}
+\sum_q\frac{n_{k+q}+\bar{n}_q}
{\epsilon_{k,\lambda}-\epsilon_{k+q,\lambda}+\omega_{q,\lambda}}
\abs{g_{q,\Lambda}}^2(1-e^2_{k,q,\lambda})
\ret & & \quad
-\sum_q\frac{1-n_{k+q}+\bar{n}_q}
{\epsilon_{k+q,\lambda}-\epsilon_{k,\lambda}+\omega_{q,\lambda}}
\abs{g_{q,\Lambda}}^2(1-e^2_{k+q,-q,\lambda})
\eeqar
The two equations (\ref{selfcon-omega}) and (\ref{selfcon-epsilon})
can be used to determine the single particle energies
self consistently. The result is correct up
to second order in $g$. The main point is that due
to the factors $(1-e^2_{k,q,\lambda})$
or $(1-e^2_{k+q,-q,\lambda})$ small energy denominators are
avoided explicitely. Without these factors
and with the renormalized energies in the second and the third term
on the right hand side replaced by the unrenormalized
energies, the
renormalization equations (\ref{selfcon-omega})
and (\ref{selfcon-epsilon}) are well known and can be
found in various textbooks on solid state theory
(see e.g. \cite{Kittel}). In the next section we show that
(\ref{selfcon-epsilon}) yields the correct renormalization of
the density of states at the Fermi surface, which is usually
calculated within the framework of Eliashberg theory.

$[\eta_\lambda,H_{I\lambda}]$ contains further terms
generating new couplings in the Hamiltonian $H_I$.
It can be written as
\beqar \label{HI}
H_{I\lambda}&=&\sum_{k,q}:g_{k,q,\lambda}c^\dagg_k c_{k+q}b^{\dagg}_q
+g^*_{k,q,\lambda}c^\dagg_{k+q} c_kb_q:
\ret & &
+\frac12\sum_{k,k^\prime,q}V_{kk^\prime q,\lambda}
:c^\dagg_{k+q} c^\dagg_{k^\prime-q} c_{k^\prime} c_{k}:
\ret & &
+\sum_q (c_q^* b^{\dagg}_qb^{\dagg}_{-q}+c_qb_qb_{-q})
\ret & &
+ \mbox{ couplings between electrons and two bosons }+O(g^3).
\eeqar
Let us calculate $V_{k,k^\prime,q}$ and $c_q$ to second order
in $g$. The other couplings can be obtained similarly.
Following (\ref{sr2}) the renormalization equation
of $V_{kk^\prime q,\lambda}$ is
\beqar
\frac{dV_{k,k^\prime ,q,\lambda}}{d\lambda}&=&
-u_{k,k^\prime ,q,\lambda}
(\eta_{k+q,-q,\lambda}g^*_{k^\prime ,-q,\lambda}
+\eta^*_{k^\prime ,-q,\lambda}g_{k+q, -q,\lambda}
+\eta_{k^\prime ,-q,q,\lambda}g^*_{k,q,\lambda}
+\eta^*_{k,q,\lambda}g_{k^\prime ,-q, q,\lambda})
\ret & &
+r_{k,k^\prime ,q,\lambda}
\frac{d \ln u_{k,k^\prime ,q,\lambda}}{d\lambda}
V_{k,k^\prime ,q,\lambda}.
\eeqar
In the first term in the right hand side
the induced electron--electron interaction
is generated due to the
elimination of the electron--phonon interaction.
The second term eliminates the induced electron--electron
interaction again. This elimination then yields higher
interactions which are of fourth or higher order
in the electron--phonon coupling. The problem is now
that we expect to obtain an attractive induced interaction
between the electrons. Such an attractive interaction
is known to be a marginal relevant operator for a fermionic system.
The elimination of such a term can cause difficulties.
Within the framework of Wegner's flow equations, this
has already been observed for the one--dimensional
problem \cite{Wegner94}. Therefore we modify the
renormalization scheme at this point. We simply choose
$u_{k,k^\prime ,q,\lambda}=1$. Then the second term
in the renormalization equation for the induced
interaction vanishes. Using the above expression for $g_{k,q,\lambda}$
and $\eta_{k,q,\lambda}$,
this equations becomes
\beqar \label{flowint}
\frac{dV_{k,k^\prime ,q,\lambda}}{d\lambda}&=&
\abs{g_{q,\Lambda}}^2
\left(\frac{e_{k^\prime ,-q,\lambda}}
{\epsilon_{k+q,\lambda}-\epsilon_{k,\lambda}+\omega_{q,\lambda}}
\frac{de_{k+q,-q,\lambda}}{d\lambda}
+\frac{e_{k+q,-q,\lambda}}
{\epsilon_{k^\prime\lambda}-\epsilon_{k^\prime-q,\lambda}
+\omega_{q,\lambda}}
\frac{de_{k^\prime ,-q,\lambda}}{d\lambda}
\right.
\ret & & \qquad
\left. +\frac{e_{k,q,\lambda}}
{\epsilon_{k^\prime -q,\lambda}-\epsilon_{k^\prime\lambda}
+\omega_{q,\lambda}}
\frac{de_{k^\prime -q,q,\lambda}}{d\lambda}
+\frac{e_{k^\prime -q,q,\lambda}}
{\epsilon_{k,\lambda}-\epsilon_{k+q,\lambda}
+\omega_{q,\lambda}}
\frac{de_{k,q,\lambda}}{d\lambda}
\right).
\eeqar
Defining
\beq
f_{k,k^\prime,q,\lambda}=\int_\lambda^\Lambda ds
e_{k^\prime -q,q,s}\frac{de_{k,q,s}}{ds},
\eeq
we can write $V_{k,k^\prime,q,\lambda}$ in the form
\beqar \label{eeinteraction}
V_{k,k^\prime, q,\lambda}=-\abs{g_{q,\Lambda}}^2
\left(\frac{f_{k,k^\prime,q,\lambda}}
{\epsilon_{k,\lambda}-\epsilon_{k+q,\lambda}+\omega_{q,\lambda}}+
\frac{f_{k^\prime -q,k+q,q,\lambda}}
{\epsilon_{k^\prime -q,\lambda}-\epsilon_{k^\prime,\lambda}+
\omega_{q,\lambda}}
\right. \ret\qquad\qquad\left. +
\frac{f_{k+q,k^\prime -q,-q,\lambda}}
{\epsilon_{k+q,\lambda}-\epsilon_{k,\lambda}+\omega_{q,\lambda}}+
\frac{f_{k^\prime,k,-q,\lambda}}
{\epsilon_{k^\prime,\lambda}-\epsilon_{k^\prime-q,\lambda}
+\omega_{q,\lambda}}
\right)+O(g^4).
\eeqar
The terms $O(g^4)$ arise again due to the fact that the
derivative of the single particle energies is of order $g^2$.
Further corrections $O(g^4)$ occur since in (\ref{flowint})
terms of $O(g^4)$ have been neglected.
(\ref{eeinteraction}) is our main result.
Note that $f_{k,k^\prime,q,\lambda}\ge 0$.
Let us discuss two interesting cases:
\begin{itemize}
\item[a)] $k^\prime=k+q$ (the diagonal part of the interaction)

We have
$f_{k,k+q,q,\lambda}=\frac12(1-e^2_{k,q,\lambda})$.
This yields
\beqar
V_{k,k+q,q,\lambda}&=&\abs{g_{q,\Lambda}}^2
\left(\frac{2\omega_{q,\lambda}}
{(\epsilon_{k+q,\lambda}-\epsilon_{k,\lambda})^2-\omega_{q,\lambda}^2}
\right. \ret & & \qquad \left.
-\frac{e^2_{k+q,-q,\lambda}}
{\epsilon_{k+q,\lambda}-\epsilon_{k,\lambda}+\omega_{q,\lambda}}
-\frac{e^2_{k,q,\lambda}}
{\epsilon_{k,\lambda}-\epsilon_{k+q,\lambda}+\omega_{q,\lambda}}
\right)
\eeqar
The first term is the well known expression already obtained
by Fr{\"o}hlich \cite{Froehlich}.
The singularity is cancelled by the two other terms.
These terms occur because the electron--phonon coupling has not
been eliminated completely.
\item[b)] $k^\prime=-k$
(the interaction of two electrons forming a Cooper pair)

Using $f_{k,-k,q,\lambda}=f_{-k,k,-q,\lambda}$ we obtain
\beq \label{cooper}
V_{k,-k, q,\lambda}=-2\abs{g_{q,\Lambda}}^2
\left(\frac{f_{k,-k,q,\lambda}}
{\epsilon_{k,\lambda}-\epsilon_{k+q,\lambda}+\omega_{q,\lambda}}
+\frac{f_{k+q,-k-q,-q,\lambda}}
{\epsilon_{k+q,\lambda}-\epsilon_{k,\lambda}+\omega_{q,\lambda}}
\right).
\eeq
In the case
$\omega_{q,\lambda}>\abs{\epsilon_{k+q,\lambda}-\epsilon_{k,\lambda}}$
both terms are negative, the
interaction is thus attractive.
If $\omega_{q,\lambda}<\abs{\epsilon_{k+q,\lambda}-\epsilon_{k,\lambda}}$
the sign of the two terms is different.
Let us consider the case
$\epsilon_{k+q,\lambda}-\epsilon_{k,\lambda}>\omega_{q,\lambda}$.
Under this condition one has
$\abs{\epsilon_{k+q,\lambda}-\epsilon_{k,\lambda}+\omega_{q,\lambda}}>
\abs{\epsilon_{k,\lambda}-\epsilon_{k+q,\lambda}+\omega_{q,\lambda}}$. For
some large cutoff $\Lambda$, both,
$\abs{\epsilon_{k+q,\Lambda}-\epsilon_{k,\Lambda}+\omega_{q,\Lambda}}$
and
$|\epsilon_{k,\Lambda}-\epsilon_{k+q,\Lambda}+\omega_{q,\Lambda}|$
are less than $\Lambda$. Lowering
the cutoff, it happens that
$\abs{\epsilon_{k+q,\lambda}-\epsilon_{k,\lambda}+\omega_{q,\lambda}}$
becomes larger than $\lambda$ whereas
$\abs{\epsilon_{k,\lambda}-\epsilon_{k+q,\lambda}+\omega_{q,\lambda}}$
is still smaller than $\lambda$.
This mean that $f_{k+q,-k-q,-q,\lambda}$ is nonzero,
whereas $f_{k,-k,q,\lambda}$ remains zero. In that case only the second
term in (\ref{cooper}) contributes. This term is negative, the
interaction is thus attractive. Similarly, if
$\epsilon_{k,\lambda}-\epsilon_{k+q,\lambda}>\omega_{q,\lambda}$,
the first term starts to contribute first and is negative.
This shows that for sufficiently large $\lambda$
and $V_{k,-k,q,\Lambda}=0$ one has
$V_{k,-k,q,\lambda}<0$. This property holds for
general $\lambda$ as well if
$e(x)$ (or $u(x)$) falls of sufficiently fast
in the interval around $x=1$. To see this, we calculate the
integral in the definition of $f_{k,-k,q,\lambda}$.
The integration from $\lambda$ to $\Lambda$ can be replaced
by an integration from $\lambda_1$ to $\Lambda$, where
$\lambda_1=\lambda$ if $e_{k,q,\lambda}e_{k+q,-q,\lambda}>0$,
otherwise $\lambda_1$ is the supremum of all $s$ with
$e_{k,q,s}e_{k+q,-q,s}=0$. In a second
step we use that  $e_{k,q,s}$, and
$e_{k+q,-q,s}$ are monotonic increasing functions of $s$.
Then the second mean value theorem for integrals yields
\beq
f_{k,-k,q,\lambda}=
\int_{\lambda_1}^{\Lambda}ds \frac{d\ln e_{k,q,s}}{ds}=
-\ln e_{k,q,\lambda_2}
\eeq
with some $\lambda_2>\lambda_1$.
This finally yields
\beq
V_{k,-k, q,\lambda}=
2\frac{\abs{g_{q,\Lambda}}^2}
{\lambda_2}
\left(\frac{\ln e_{k+q,-q,\lambda_2}}{x_{k+q,-q,\lambda_2}}
+\frac{\ln e_{k,q,\lambda_2}}{x_{k,q,\lambda_2}}\right)
\eeq
where
$x_{k,q,\lambda}=(\epsilon_{k,\lambda}-\epsilon_{k+q,\lambda}
+\omega_{q,\lambda})/\lambda$.
Whenever $\abs{\ln e(x)}/ \abs{x}$ is a monotonously increasing
function of $\abs{x}$,
the right hand side is negative.
This condition can easily be satisfied. Let us write
$e(x)$ in the form $e(x)=\exp(-\abs{x} h(\abs{x}))$.
Then the condition is satisfied if $h(x)$
increases monotonously as a function of $x$.
This means that the functions $e(x)$ and $u(x)$ have
to decay sufficiently fast.
\end{itemize}

To obtain an explicit expression for the interaction within
a Cooper pair, one has to choose a specific form of
$e(x)$. If we assume that $e(x)$ drops rapidly from 1 to 0, we can take
for simplicity $e(x)=\theta(1-x)\theta(1+x)$.
This choice of a step function is useful for doing
analytical calculations, but as mentioned above it can lead to pathologies
in higher orders in $g$. With this choice
of $e(x)$ we obtain
\beq
f_{k,-k,q,\lambda}=\theta(\epsilon_{k,\lambda}-\epsilon_{k+q,\lambda})
\theta(\abs{\epsilon_{k,\lambda}-\epsilon_{k+q,\lambda}+\omega_{q,\lambda}}
-\lambda).
\eeq
This yields
\beq \label{rc}
V_{k,-k,q,\lambda}=-\frac{2\abs{g_{q,\Lambda}}^2}
{\abs{\epsilon_{k+q,\lambda}-\epsilon_{k,\lambda}}+\omega_{q,\lambda}}
\theta(\abs{\epsilon_{k+q,\lambda}-\epsilon_{k,\lambda}}+\omega_{q,\lambda}
-\lambda).
\eeq
The form (\ref{rc}) of the phonon--induced electron--electron interaction
is similar to the result obtained by Lenz et al \cite{Lenz96},
see (\ref{LWint}), except for the cutoff function.
It differs from the result by Fr{\"o}hlich
(\ref{Fint}). The three expressions agree for
$\abs{\epsilon_{k+q,\lambda}-\epsilon_{k,\lambda}}=0$.

The coupling $c_{q,\lambda}$ can be calculated likewise.
We obtain
\beqar
\frac{dc_{q,\lambda}}{d\lambda}&=&-\frac12u(x_{q,\lambda})
\sum_kn_k
(\eta_{k,q,\lambda}g_{k+q,-q,\lambda}
+\eta_{k+q,-q,\lambda}g_{k,q,\lambda}
+\eta_{k-q,\lambda}g_{k-q,q,\lambda}
+\eta_{k-q,q,\lambda}g_{k-q,\lambda})
\ret & &
+r(x_{q,\lambda})\frac{d\ln u(x_{q,\lambda})}{d\lambda}c_{q,\lambda}
\eeqar
where
\beq
x_{q,\lambda}=\frac{2\omega_{q,\lambda}}
{\lambda}.
\eeq
With a similar ansatz as above,
\beq
c_{q,\lambda}=e(x_{q,\lambda})\tilde{c}_{q,\lambda}
\eeq
we have
\beqar
\frac{d\tilde{c}_{q,\lambda}}{d\lambda}&=&-\frac12\exp(-r(x_{q,\lambda}))
\sum_kn_k
(\eta_{k,q,\lambda}g_{k+q,-q,\lambda}
+\eta_{k+q,-q,\lambda}g_{k,q,\lambda}
\ret & & \qquad\qquad\qquad\qquad\qquad
+\eta_{k,-q,\lambda}g_{k-q,q,\lambda}
+\eta_{k-q,q,\lambda}g_{k,-q,\lambda}).
\eeqar
$c_{q,\lambda}$ can be calculated  using the same
approximations as above. We obtain
\beqar
\tilde{c}_{q,\lambda}=\frac12\abs{g_{q,\Lambda}^2}
\sum_k n_k\left(
\frac{f_{k,q,\lambda}}{\epsilon_{k,\lambda}-\epsilon_{k+q,\lambda}
+\omega_{q,\lambda}}+
\frac{f_{k+q,-q,\lambda}}{\epsilon_{k+q,\lambda}-\epsilon_{k,\lambda}
+\omega_{q,\lambda}}\right.
\ret \qquad\qquad\left. +
\frac{f_{k,-q,\lambda}}{\epsilon_{k,\lambda}-\epsilon_{k-q,\lambda}
+\omega_{q,\lambda}}
+\frac{f_{k-q,q,\lambda}}{\epsilon_{k-q,\lambda}-\epsilon_{k,\lambda}
+\omega_{q,\lambda}}\right)
\eeqar
where
\beq
f_{k,q,\lambda}=\int_\lambda^\Lambda ds
\exp(-r(x_{q,s}))e_{k+q,-q,s}\frac{de_{k,q,s}}{ds}.
\eeq
In the final Hamiltonian these couplings can be treated
perturbatively. They yield a renormalization of the
phonon frequencies in fourth order in $g_{q,\Lambda}$.
Similarly, the additional terms in the Hamiltonian $H_I$
which contain couplings of an electron to two phonons
yield an additional contribution to the
electron--electron interaction of fourth order in $g$.

\Section{Some explicit results for the Einstein model}
The above results are very general and can be applied to
any type of electron--phonon interaction, even to any
type of interaction between electrons and other bosonic degrees
of freedom. In this section we apply our results to
the simple Einstein model of phonons. Furthermore
we will assume a constant density of states for the
electrons near the Fermi surface, and we neglect
renormalization effects of the phonon energies. The motivation is
to show the reader how well known results can be obtained
using the above procedure. We will also show what are
the main effects of the new phonon--induced
electron--electron interaction (\ref{rc}) on the
critical temperature.

In the Einstein model one takes phonons with
$\omega_q=\omega_0$ and $g_{q,\Lambda}=g_0V^{-1/2}$.
These assumptions yield considerable simplifications in the
above formulae. Let us first study the renormalization of the
electronic energies $\epsilon_{k,\lambda}$. From (\ref{selfcon-epsilon})
we obtain
\beqar \label{epsilon-E1}
\epsilon_{k,\lambda}&=&\epsilon_{k,\Lambda}
+g_0^2V^{-1}\sum_{k^\prime}\frac{n_{k^\prime}}
{\epsilon_{k,\lambda}-\epsilon_{k^\prime,\lambda}+\omega_0}
\theta(\abs{\epsilon_{k,\lambda}-\epsilon_{k^\prime,\lambda}+\omega_0}
-\lambda)
\ret & & \quad
-g_0^2V^{-1}\sum_{k^\prime}\frac{1-n_{k^\prime}}
{\epsilon_{k^\prime,\lambda}-\epsilon_{k,\lambda}+\omega_0}
\theta(\abs{\epsilon_{k^\prime,\lambda}-\epsilon_{k,\lambda}+\omega_0}
-\lambda),
\eeqar
at zero temperature, where the phonon occupation
number $\bar{n}_q= 0$.
As usual we replace the summation over $k^\prime$ by an integration
over $\epsilon_{k^\prime}$. Furthermore we assume that the density
of states is constant, we denote it by $N(0)$.
The integral over $\epsilon_{k^\prime}$ extends over the whole
band and we assume that the band width $D$ is large compared to
$\omega_0$ and $\lambda$. The
renormalization of band energies $\epsilon_{k,\lambda}$
with $\abs{\epsilon_{k,\lambda}}<\omega_0$ is then
given by
\beq \label{epsrenormE1}
\epsilon_{k,\lambda}=\epsilon_{k,\Lambda}
+g_0^2N(0)\ln\frac{\omega_0-\epsilon_{k,\lambda}
+\max(0,\lambda+\epsilon_{k,\lambda}-\omega_0)}
{\omega_0+\epsilon_{k,\lambda}
+\max(0,\lambda-\epsilon_{k,\lambda}-\omega_0)}.
\eeq
For a cutoff $\lambda$ that is smaller than the phonon frequency $\omega_0$
and for $\epsilon_k\ll\omega_0$ this yields the well
known expression (see e.g. \cite{Scalapino})
\beq \label{epsrenormE2}
\epsilon_{k,\lambda}=\frac{\epsilon_{k,\Lambda}}{1+2g_0^2N(0)/\omega_0}.
\eeq
The last expression can be derived in the framework of Eliashberg theory,
which is valid under the same conditions as used for the derivation
of (\ref{epsrenormE1}), namely $\omega_0\ll D$.
An important point in this discussion is that the cutoff
$\lambda$ should be somewhat less but of the order of $\omega_0$.
With this choice properties of electrons close to the Fermi
surface are described correctly. If the cutoff is larger, a
part of the electron--phonon interaction that is relevant for
the electronic behaviour near the Fermi surface has not been
eliminated. A lower cutoff does not affect the
single particle electronic properties
near the Fermi surface.

Let us now come to the discussion of the phonon--induced
electron--electron interaction. The expression
(\ref{rc}) for the interaction of two electrons
forming a Cooper pair contains the factor
$\theta(\abs{\epsilon_{k+q,\lambda}-\epsilon_{k,\lambda}}+\omega_{0}
-\lambda)$.
This factor arises due the fact that for a given $\lambda$
only a part of the electron--phonon interaction has been eliminated.
If we choose $\lambda$ to be less than $\omega_0$, this
factor is unity. This is similar to the behaviour of the single
particle energies. This part of the interaction then no longer depends
on the cutoff $\lambda$.

The energy gap and the critical temperature can be estimated if we
treat the renormalized Hamiltonian as in BCS-theory.
This leads to the usual BCS equation for the energy gap, where
the electron--electron interaction (\ref{rc}) has to be taken,
\beq \label{BCS}
\Delta_k=\sum_q\frac{V_{k,-k,q,\lambda}\Delta_{k+q}}
{2\sqrt{\epsilon_{k+q,\lambda}^2+\Delta_{k+q}^2}}\tanh
\left(\frac{\beta}2\sqrt{\epsilon_{k+q,\lambda}^2+\Delta_{k+q}^2}\right).
\eeq
In the isotropic case
we can replace the sum over $q$ by an integral over $\epsilon_{k+q}$.
Introducing  $\Delta(\epsilon_k)=\Delta_k$ and using (\ref{rc}) we obtain
\beq
\Delta(\epsilon)=2g_0^2\int
d\epsilon^\prime \rho(\epsilon^\prime)
\frac{1}
{\abs{\epsilon-\epsilon^\prime}+\omega_0}
\frac{\Delta(\epsilon^\prime)}
{2\sqrt{\epsilon^{\prime\,2}+\Delta(\epsilon^\prime)^2}}
\tanh\left(\frac{\beta}2\sqrt{\epsilon^{\prime\,2}+\Delta(\epsilon^\prime)^2}
\right).
\eeq
Here $\rho(\epsilon^\prime)$ is the renormalized density of states
since (\ref{BCS}) has been derived from the renormalized Hamiltonian
and therefore we have to use renormalized
energies. It can be approximated by a constant renormalized
density of states
$\rho(0)=N(0)/(1+2g_0^2N(0)/\omega_0)$. Using dimensionless
quantities
$x=\epsilon/\omega_0$, $x^\prime=\epsilon^\prime/\omega_0$,
$\tau=(\beta\omega_0)^{-1}$,
$\bar{\Delta}(x)=\Delta(\epsilon)/\omega_0$
we obtain
\beq \label{Delta-sr}
\bar{\Delta}(x)=\frac{2g_0^2N(0)/\omega_0}{1+2g_0^2N(0)/\omega_0}
\int dx^\prime
\frac{\bar{\Delta}(x^\prime)}{\abs{x-x^\prime}+1}
\frac{\tanh(\sqrt{x^{\prime\, 2}+\bar{\Delta}(x^{\prime})^2}/\tau)}
{\sqrt{x^{\prime\, 2}+\bar{\Delta}(x^{\prime})^2}}.
\eeq
This equation can be compared to the usual BCS equation with
a constant interaction in an energy interval given by $\omega_0$,
\beq \label{Delta-BCS}
1=2g_0^2N(0)/\omega_0\int_{\abs{x^\prime}<1} dx^\prime
\frac{\tanh(\sqrt{x^{\prime\, 2}+\bar{\Delta}^2}/\tau)}
{\sqrt{x^{\prime\, 2}+\bar{\Delta}^2}}
\eeq
The main difference between our result (\ref{Delta-sr}) and the BCS result
is the renormalization of the density of states $N(0)$. It is well
known (see e.g. the discussion by Allen et al, \cite{Allen82, Allen90})
that this renormalization is important when the coupling
$2g_0^2N(0)/\omega_0$ is not small. Another important point is the
different way the cutoff is introduced. In BCS theory one
has to introduce a cutoff by hand, and this is usually done by
restricting the electronic energies to an energy interval
of width $2\omega_0$ around the Fermi surface.
The main motivation for this is that an interaction of
the Fr{\"o}hlich--type (\ref{Fint}) contains a singularity
and becomes repulsive for larger energies. In our case
we use a renormalization procedure which leads automatically
to a cutoff $\omega_0$ due to the factor
$(\abs{x-x^\prime}+1)$ in the denominator.

(\ref{Delta-sr}) cannot be solved analytically, but it is easy to solve
it numerically. In Fig. 1 we show the solution for $\bar{\Delta}(x)$
for $2g_0^2N(0)/\omega_0=1/3$ and
$2g_0^2N(0)/\omega_0=1.4$ for different temperatures.
One clearly sees that the gap becomes
small for large $x$ and decreases with increasing temperature.
The scaled curves in Fig. 1 show that
the scaling relation
\beq \label{Deltascaling}
\Delta(\omega,T)=\Delta(\omega,0)\frac{\Delta(0,T)}{\Delta(0,0)}
\eeq
is satisfied within the numerical accuracy. We have checked this scaling
law numerically for values of $2g_0^2N(0)/\omega_0$ up to 2.5. The
scaling law has been found by Scalapino et al \cite{SWS}
(see also \cite{Scalapino}) for weak coupling superconductors,
whereas in the strong coupling case deviations occurred. But their
calculations have been done for more complicated phonon
spectra so that a direct comparison with our results is not
possible. Scalapino \cite{Scalapino} also mentions that even for strong
coupling the reduced energy gap $\Delta(0,T)/\Delta(0,0)$
as a function of $T/T_c$ follows very closely the BCS
curve for this quantity. We have plotted the reduced energy gap
in Fig. 2, again for $2g_0^2N(0)/\omega_0=1/3$
and $2g_0^2N(0)/\omega_0=1.4$. For other values of
the coupling strength the curve is the same, it agrees with the BCS
curve. The data shown in Fig. 2a show that for small
values of the coupling the ratio $2\Delta(0,0)/T_c=3.53$
as predicted by BCS--theory. For $2g_0^2N(0)=1.4$ we obtain
$\Delta(0,0)=0.298\omega_0$ and $T_c=0.161\omega_0$. In that
case the ratio $2\Delta(0,0)/T_c=3.70$. This tendency to
larger values of $2\Delta(0,0)/T_c$ for strongly coupled
systems in well known.
The behaviour of $\Delta(\omega,T)$ for other values of
$2g_0^2N(0)/\omega_0$ is similar. The main difference is the value
of $\bar{\Delta}(0)$, which shows a strong dependence on
$2g_0^2N(0)/\omega_0$ as expected.

An important point is that the
energy gap $\bar{\Delta}(0)$ is about an order of magnitude
(for small couplings $2g_0^2N(0)/\omega_0\lessapprox 0.5$ a factor
of 5) smaller
than the value one obtains using (\ref{Delta-BCS}).
As a consequence the calculated critical temperature
is smaller by roughly the same factor. We show this
general feature in Fig. 3, where $T_c$ is plotted
as a function of $2g_0^2N(0)/\omega_0$ together with the BCS--result.
A comparison with the
value of the critical temperature
calculated with the interaction (\ref{LWint})
from Lenz and Wegner \cite{Lenz96} is not so easy, since in
their paper the renormalization of the electronic single
particle energies was not taken into account. But the flow
equation for $\epsilon_k$ in their paper is similar to what
we obtained, so that one should expect a similar renormalization
of the single particle energies. Therefore we put this
in by hand. The values for the gap calculated with
the same assumptions as above show a behaviour that is very
similar to
our results. Only for larger values of $2g_0^2N(0)/\omega_0$ deviations
occur. For $2g_0^2N(0)/\omega_0\gtrapprox 1.2$ the gap for
the interaction (\ref{LWint}) lies below our curve.
Furthermore it is interesting to compare our result with
results for $T_c$ obtained in the framework of Eliashberg theory.
One famous result is the McMillan--Dynes equation
\cite{McMillan, Dynes}, which
has in our case the form
\beq
T_c=\frac{\omega_0}{1.2}\exp\left(-\frac{1.04(1+2g_0^2N(0)/\omega_0)}
{2g_0^2N(0)/\omega_0}\right).
\eeq
This curve has also been plotted in Fig. 3. The plot shows that
for small and intermediate coupling the agreement is very good,
whereas for strong coupling ($2g_0^2N(0)/\omega_0\gtrapprox 1.2$) deviations
occur. It is well known that in the strong coupling case the McMillan--Dynes
equation is not valid. But in the region where the McMillan--Dynes
equation is applicable, we obtain a good agreement.
These results clearly show that a renormalization procedure
for Hamiltonians like similarity renormalization or
flow equations treats the different energy scales in the problem
in a satisfactory way.

\Section{Discussion of the results}
Starting from an initial Hamiltonian with an electron--phonon
interaction, we calculated an effective Hamiltonian to second
order in the electron--phonon coupling. Due to the
special structure of the similarity renormalization scheme,
the effective Hamiltonian has a band--diagonal
structure. It contains still a small part of the
electron--phonon interaction, in addition an
effective electron--electron interactions and other
couplings. The renormalized single particle energies
and the induced interactions are free of divergencies.
The actual values of the renormalized quantities
depend on the special choice of the infinitesimal
unitary transformation, i.e. on the cutoff $\lambda$ and
on the special choice of the
function $u(x)$. On the other hand it
is clear that measurable quantities like
expectation values of observables should not depend on
$\lambda$ or $u(x)$. This is not a contradiction, since
the unitary transformation applied to the Hamiltonian has to
be applied to the observables as well.
We have only calculated the effective Hamiltonian,
thus we are not able to check this explicitely. But
we have shown that the structure of the Hamiltonian does
not depend on $u(x)$.
The renormalization of the single particle energies
shows no essential dependence on $u(x)$ if the cutoff
$\lambda$ is sufficiently small.
With some additional weak condition
on $u(x)$ ($u(x)$ has to decay sufficiently fast), {\em
the induced interaction between two electrons forming
a Cooper pair is attractive in the whole parameter space}.

Comparing our results for the induced electron--electron
interaction with Fr{\"o}hlich's result
(\ref{Fint}) shows that the effective interaction obtained by
Fr{\"o}hlich has not only the problem that it contains
a divergency, even the sign of the interaction is wrong
in some part of the parameter space.
In contrast
the effective interaction obtained by Lenz and Wegner
shows the correct behaviour. Another difference
is that in the perturbative expression of Fr{\"o}hlich the
unrenormalized single particle energies enter.
As a consequence, the renormalization of the electronic density of
states, which has an important effect when one calculates $T_c$,
is not taken into account. As already mentioned in the
introduction, a similar behaviour has been observed
in the case of induced interactions for other problems as well
\cite{KM}.
The fact that the induced interaction in the general form
(\ref{cooper}) is always attractive can be traced back to
the correct treatment of the different energy scales. Whenever
large energy differences are treated first, the interaction
(\ref{cooper}) is attractive. If, as in the treatment
by Fr{\"o}hlich \cite{Froehlich} or by Bardeen and Pines
\cite{BP} all energy scales are treated at the same time,
$f_{k,-k,q,\lambda}$ does not depend on $k$ and $q$.
In that case (\ref{cooper})
yields a result similar to the one obtained by Fr{\"o}hlich.

The discussion of the simple model in section 4 has shown,
that the similarity renormalization scheme can be
successfully applied to the electron--phonon problem.
It yields the correct energy scale for
single particle excitations and an effective interaction
that yields reasonable values for the energy gap or for $T_c$.
But several problems have been left open. The reason
was that the simple example was only presented
to illustrate how the method works. Applying the
similarity renormalization scheme to a more realistic model
is possible and in that case a more detailed study
should be done. For instance, the renormalization of
the phonon frequencies has not been take into account.
Furthermore, the renormalization procedure has been
performed only to second order in the electron--phonon coupling.
The similarity renormalization used here is a perturbative
renormalization procedure. Higher corrections to
the electron--electron interaction can be calculated
systematically. The non--linear electron--phonon
interaction in (\ref{HI}) e.g. yields a contribution
to the induced electron--electron interaction in
fourth order.
A detailed discussion of such higher terms may be important, since
the attractive interaction is known to be a marginal relevant
operator in a problem of interacting fermions.
Furthermore it is possible to
include the initial Coulomb repulsion in the Hamiltonian.
It will be modified by the electron--phonon interaction
as well, but it does not introduce a new energy scale to
the problem. In this way it should be possible to obtain
a more complete expression for the electron--electron
interaction. We leave these problems for future work.

Let us now compare the similarity renormalization scheme
with Wegner's flow equations for Hamiltonians.
A common feature of both approaches is that they
work in a Hamiltonian framework. As a consequence,
we will never be able to obtain effective interactions
containing retardation effects.
The Hamiltonian is always a hermitian operator
and induced interactions never depend
on frequency. This is in contrast to methods
using path integrals. In a path integral formulation of
the present problem one is able to integrate
out the phonons completely. The effective theory
then contains a time--dependent
or frequency--dependent electron--electron
interaction. The interaction obtained in this
way has nothing to do with the effective interaction in
our approach. But this does not mean that one is not
able to describe the decay of metastable states or
the lifetime of some bound state in a Hamiltonian framework.
In a recent paper on dissipative quantum systems this point
has been discussed in great detail \cite{KM2}.

Both approaches, flow equations and the similarity renormalization scheme
use a continuous unitary transformation
to construct an effective Hamiltonian from a given
initial Hamiltonian.
This is typical for renormalization procedures.
Already in the old formulation of renormalization
Wilson \cite{Wilson65} used a sequence of unitary transformations
to calculate an effective Hamiltonian with a small cutoff.
But the goal is different in both
cases. In the flow equation approach, the Hamiltonian
is transformed into a block--diagonal form.
In the present case of the electron--phonon
coupling the number of phonons is conserved in
each block. But the Hamiltonian may still contain off--diagonal
elements between two states with a large energy difference.
In impurity models as in the Anderson
impurity model \cite{KM}, or in dissipative quantum systems,
where a small system is coupled to a bath \cite{KM2},
this does not happen, but there is no way to exclude
such matrix elements in general. On the other hand,
Wegner's flow equations can be used to diagonalize
a given Hamiltonian approximately \cite{Wegner94}. If in that case
one performs the integration of the flow equations only
to a finite value of the flow parameter, one obtains a Hamiltonian with
a band--diagonal structure. The far off--diagonal
matrix elements do not vanish, but they become
exponentially small.

When a Hamiltonian is transformed into a block--diagonal
form, some interactions like the electron--phonon
coupling in our case are eliminated completely.
It is even possible to choose the continuous
unitary transformation in such a way that some of the
possible new induced interactions do not occur.
This never happens in the similarity renormalization scheme.
It has been designed to obtain an effective
Hamiltonian with a band diagonal structure.
As a consequence, the final Hamiltonian will contain
any possible induced interaction. An example is the coupling
$c_{q,\lambda}$ that occurs in our effective Hamiltonian.
It has been avoided explicitely in the paper by Lenz and Wegner.
On the other hand, all the induced couplings in the
similarity renormalization scheme are free of divergencies.
As a consequence, a systematic expansion of matrix elements
in a coupling constant is always possible and well defined.
The induced interactions one obtains in the flow
equation approach are less singular than similar
expression obtained perturbatively, but they may
still contain some weak divergencies.  An expansion in a
small coupling constant is therefore not automatically
well defined.
But due to the simpler structure of the final Hamiltonian,
it is much easier to calculate expectation values
of observables or dynamical properties in the flow
equation approach. Therefore, depending on the
problem one wants to treat, one or the other approach
will have advantages.

\vskip1cm
\subsubsection*{Acknowledgement}

I thank F.~Wegner, S.~K.~Kehrein and P.~Lenz
for many helpful discussions. Further I thank B. M{\"u}hlschlegel
for some helpful remarks and suggestions.

\pagebreak

\pagebreak \thispagestyle{empty}
\setcounter{section}{1}

\noindent
\subsection*{Figure captions}

\renewcommand{\labelenumi}{{Fig.} \arabic{enumi}.}
\begin{enumerate}
\item The energy gap as a function of $\omega$ for
$2g_0^2N(0)/\omega_0=1/3$ (a) and $2g_0^2N(0)/\omega_0=1.4$ (b) and
different values of $T$. In Fig. 1a the temperatures
are $T=0$ (solid line), $T=0.0075\omega_0$ (dashed line),
and $T=0.01\omega_0$ (long dashed line). In Fig. 2a the temperatures
are $T=0$ (solid line), $T=0.12\omega_0$ (dashed line),
and $T=0.15\omega_0$ (long dashed line).The scaled curves
(using (\ref{Deltascaling})) lie on top of the solid line.

\item The gap at $\omega=0$ as a function of temperature
for $2g_0^2N(0)/\omega_0=1/3$ (a) and $2g_0^2N(0)/\omega_0=1.4$ (b).

\item $T_c$ as a function of the coupling $2g_0^2N(0)/\omega_0$.
The solid line shows the result from
(\ref{Delta-sr}). The dashed line is the result
obtained with the interaction of Lenz and Wegner, as explained
in the text. The long dashed line shows the result of BCS--theory,
obtained from (\ref{Delta-BCS}), and the dashed-dotted line is
a plot of the McMillan--Dynes formula.
\end{enumerate}

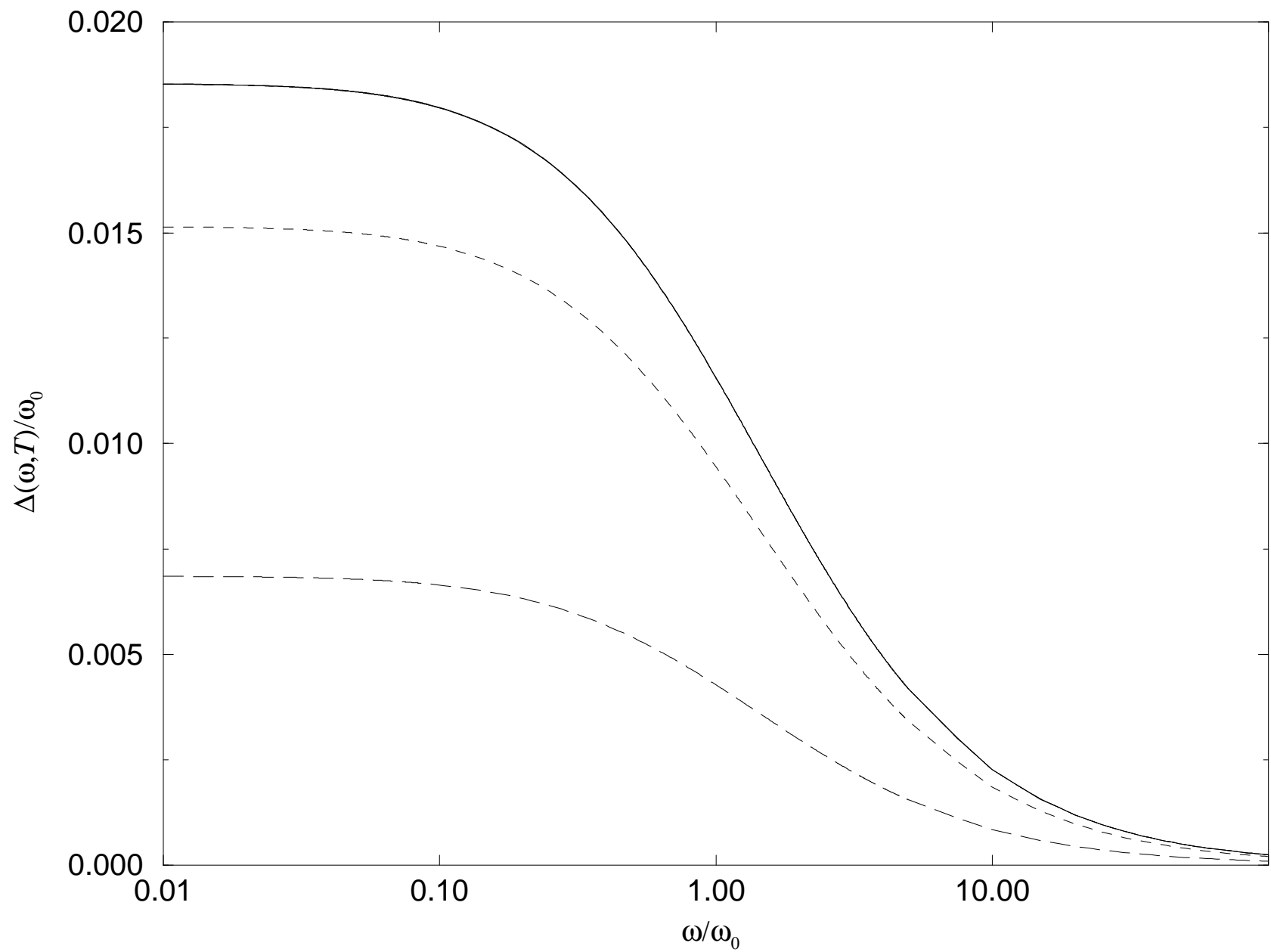
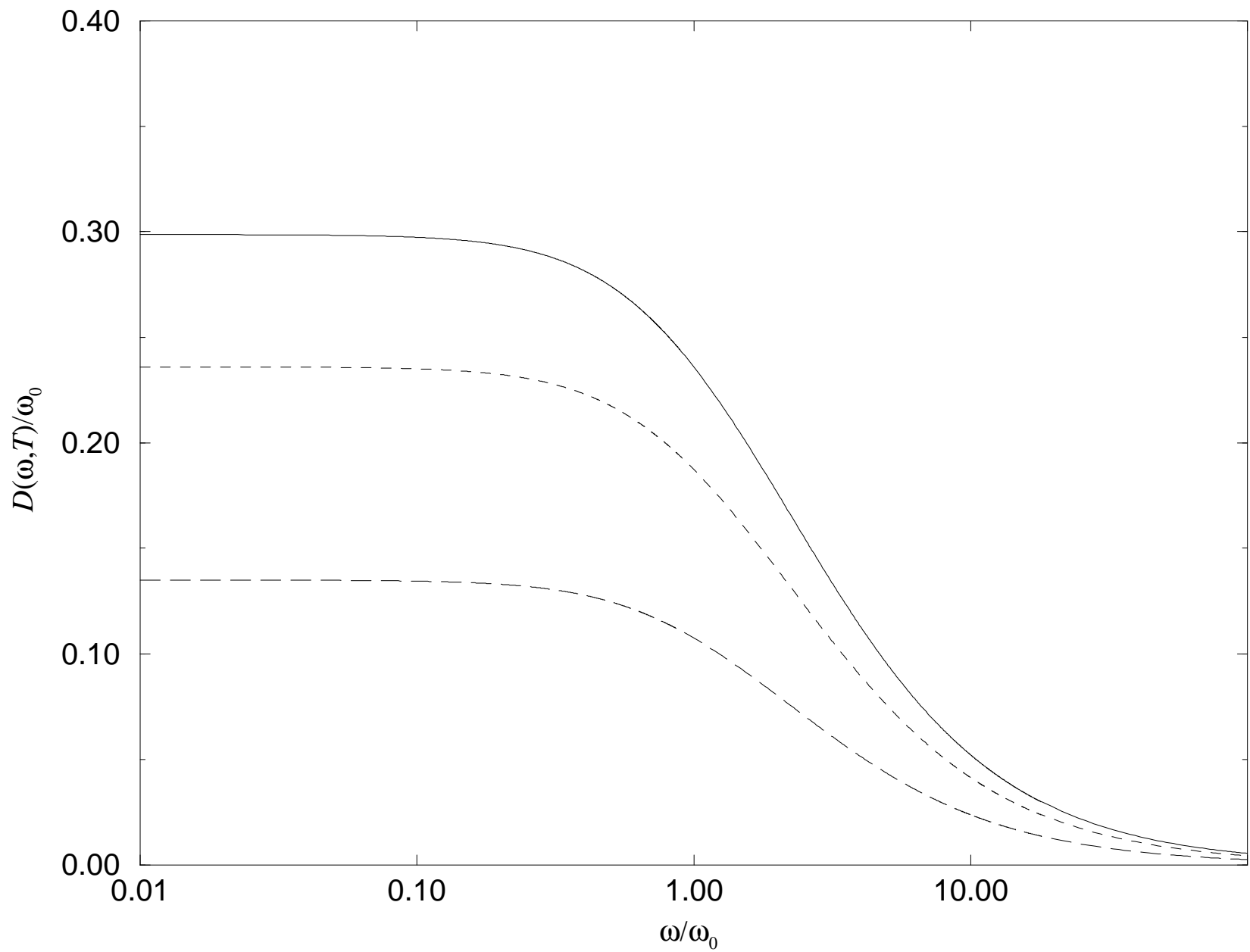
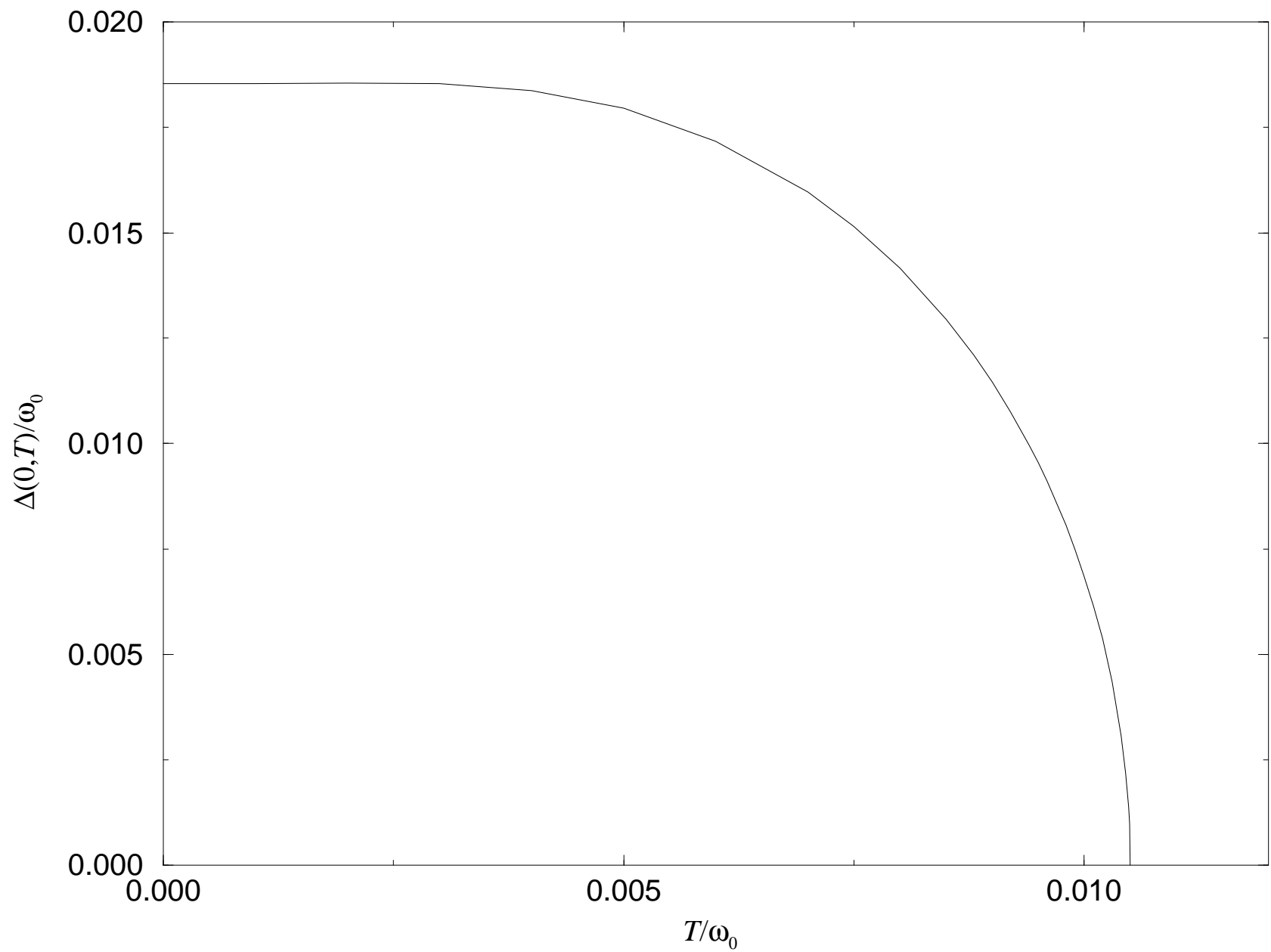
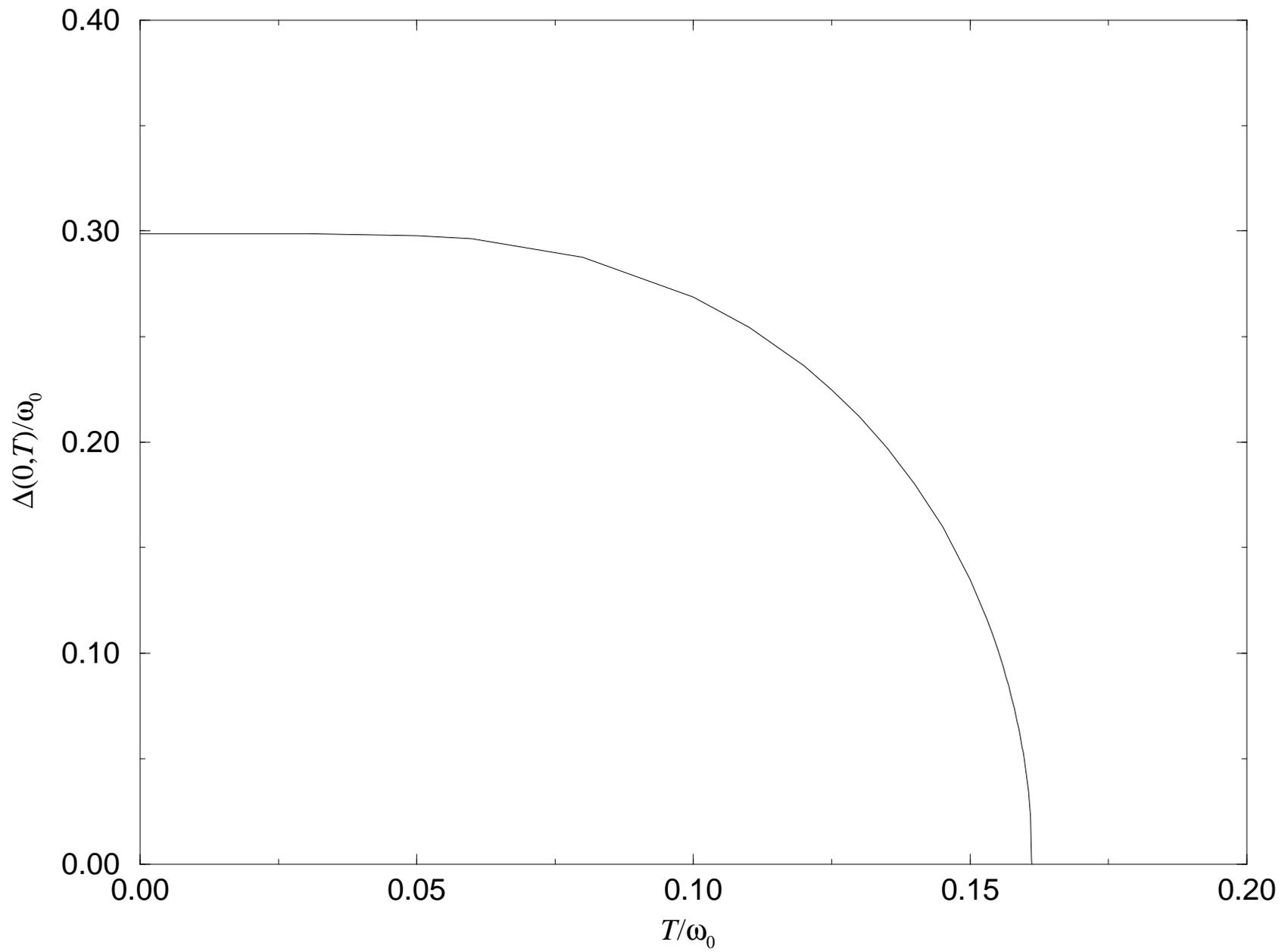
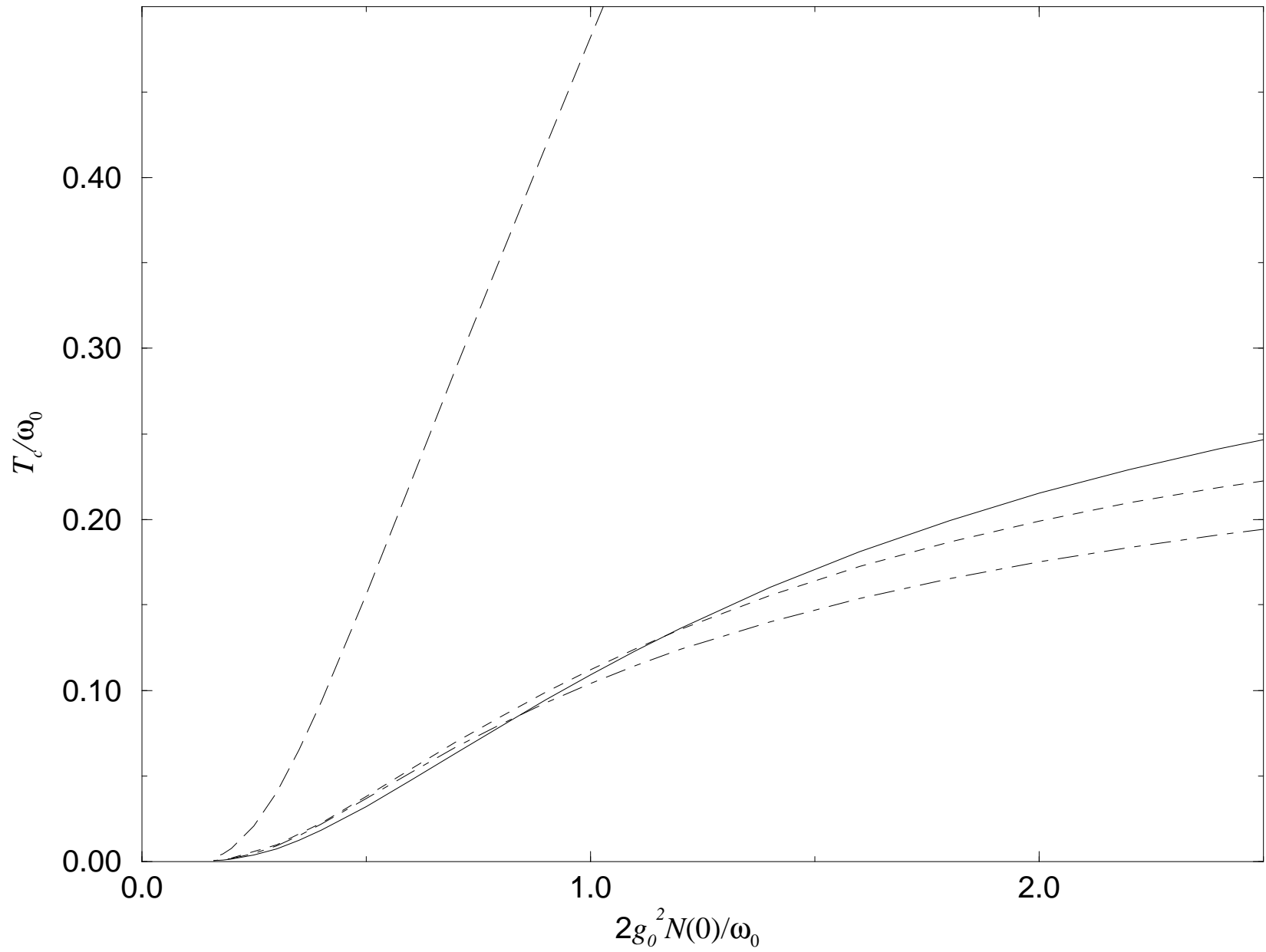
\end{document}